# Thermal expansion and polyamorphism of $N_2 - C_{60}$ solutions

V.G. Manzhelii[1], A.V. Dolbin[1], V.B. Esel'son[1], V.G. Gavrilko[1],

G.E. Gadd[2], S. Moricca[2], D. Cassidy[2], B. Sundqvist[4].

[1] Institute for Low Temperature Physics & Engineering NASU, Kharkov 61103, Ukraine

[2] Australian Nuclear Science & Technology Organisation, NSW 2234, Australia

[3] Department of Physics, Umea University, SE - 901 87 Umea, Sweden

Electronic address: dolbin@ilt.kharkov.ua

## Abstract

The linear thermal expansion coefficients α (T) of $N_2 - C_{60}$ solutions with 9.9% and 100% of the $C_{60}$ lattice interstitials filled with $N_2$ have been investigated in the interval 2.2 – 24 K. The dependence α (T) has a hysteresis suggesting co-existence of two types of orientational glasses in these solutions. The features of the glasses are compared. The characteristic times of phase transformations in the solutions and reorientation of $C_{60}$ molecules have been estimated.

## Introduction

At room temperature fullerite $C_{60}$ has a FCC lattice. The molecules of the lattice are orientationally disordered and perform a weakly hindered rotation. When the temperature decreases, the rotation slows down gradually and at 260 K there occurs a transition into a partially orientationally ordered phase. The phase has a lattice in which the molecules rotate about the three axes directed along the main cube diagonals. Below 90 K the molecular orientations are frozen completely and an orientational glass appears. The crystal lattice of $C_{60}$ has quite large interstitial cavities which can be occupied by molecules of other substances if they have dimensions permitting them to penetrate inside the cavities. In the $C_{60}$ lattice each molecule has one octahedral and two tetrahedral cavities.

Dilatometric investigations [1 – 3] on single and polycrystal samples of high-purity (99.99 wt. %) $C_{60}$ revealed a negative thermal expansion coefficient whose magnitude was very high. Arguments were advanced that attribute this unexpected effect to the tunnel rotation of the $C_{60}$ molecules. A theoretical model was proposed [4, 5], which explains qualitatively the possibility of tunnel rotation of $C_{60}$ molecules.

The influence of admixtures upon the thermal expansion of $C_{60}$ was investigated in the interval 2 – 20 K in [3, 6 – 8]. The admixtures added to $C_{60}$ were gases consisting of spherically symmetrical atoms (He, Ne, Ar, Kr, Xe) and molecules $H_2$ and $D_2$. The thermal expansion of these solutions was described as a sum of positive and negative contributions. The positive contribution was made by the low-frequency excitations of the $C_{60}$ lattice (phonons and librons); the negative contribution came from the tunnel reorientation of the $C_{60}$ molecules. The characteristic times of the contributions and their temperature dependences were estimated. The magnitudes, temperature dependences and the characteristic times of negative contributions exhibited a strong dependence on the type of the dissolved gas. In all experiments the characteristic times of the negative contributions were appreciably larger than those of the positive contributions, i.e. the process of reorientation of $C_{60}$ molecules slower than the thermalization of the phonon and libron subsystems of the $C_{60}$ lattice.

2In contrast to particles of other gases, He atoms and $H_2$, $D_2$ molecules were found to occupy not only the octahedral cavities of the $C_{60}$ lattice, but the smaller tetrahedral ones as well. The presence of He atoms in the tetrahedral cavities of $C_{60}$ was clearly demonstrated in [9, 10].

It is found that some gases dissolved in $C_{60}$ can cause a first-order phase transition in the orientational glass [3, 8]. This phenomenon, named polyamorphism [11, 12], occurs when the diameters of the atoms (molecules) introduced into $C_{60}$ exceed the effective dimensions of the interstitial cavities of $C_{60}$. An X-ray investigation [13] of the phase transformation in the orientational glass of the Xe – $C_{60}$ solution covered the whole interval of temperatures where the orientational glass exists. The thermodynamic aspects of polyamorphous transformations in $C_{60}$ saturated with gases were considered in [8].

To continue the above studies [3, 6 – 8], it was expedient to investigate $C_{60}$ solutions with gases having nonspherical molecules. This paper describes the thermal expansion of the $N_2$ – $C_{60}$ solution in the interval 2.2 – 24 K at high and low concentrations of the dissolved gas. The use of updated experimental equipment has enabled measurements that were impossible in the previous experiments [3, 6 – 8]. In particular, it is now possible to measure quite accurately the amount of absorbed gas at low concentrations in $C_{60}$ and the temperature dependence of the characteristic time of phase transformations in the orientational glass.

## Experimental technique and results.

The experiment at low $N_2$ concentrations was made on the $C_{60}$ sample that had been used in [8] to investigate the thermal expansion of the $H_2$ – $C_{60}$ and $D_2$ – $C_{60}$ systems. Before saturating it with $N_2$, the sample had to be cleaned from the gas impurities, which was done by keeping it at 300 – 400° C for ten days under the condition of dynamic evacuation (1· $10^{-3}$ Hg mm). Then, the linear thermal expansion coefficient was measured on the cleaned sample. The coefficient obtained in this control measurement was positive in the whole interval of temperatures 2.2 – 24 K (Fig. 1, curve 3). Curve 3 coincides within the experimental error with the results for pure $C_{60}$ (sample IV in [3]). The temperature dependence of the thermal expansion coefficient of the impurity-free $C_{60}$ sample exhibited no hysteresis either on heating or on cooling. The negative contribution to α (T) was absent in the whole range of temperatures.

To prepare a $N_2$ – $C_{60}$ solution with a low $N_2$ concentration, the measuring cell with the sample was filled with $N_2$ up to the pressure 760 Hg mm. The filling was carried out at room temperature. The sample was kept in the measuring cell under these conditions for 20 days. The $N_2$ concentration in $C_{60}$ equal to the ratio between the number of $N_2$ molecules and the number of $C_{60}$ molecules was 9.9%. Our method of determining the gas concentration in $C_{60}$ will be described in a special publication.

The measuring cell with the sample was then cooled slowly (for eight hours) down to 65 K. On reaching this temperature, the cell with the sample was evacuated to pressure no worse than 1· $10^{-5}$ Hg mm and the sample was further cooled for 50 minutes to the boiling point of liquid helium at which the sample was kept for 4 hours before the dilatometric investigation was started.

The function α (T) for the $N_2$- doped $C_{60}$ sample is shown in Fig. 1. All the curves are averaged over the data of several experimental series. It is seen that the heating and subsequent cooling of the $N_2$ – $C_{60}$ sample at temperatures above 4.5 K cause a hysteresis in α (T). Below 4.5 K the α (T) values taken on heating and cooling coincided.

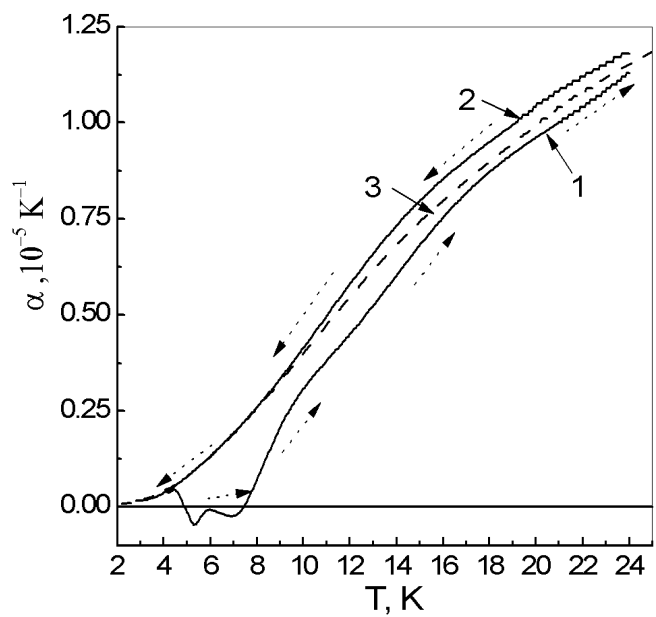

Fig. 1. Temperature dependence of the linear thermal expansion coefficient of pure $C_{60}$ and $C_{60}$ with 9.9% $N_2$
1 - $N_2 – C_{60}$ on heating, 2 - $N_2 – C_{60}$ on cooling, 3 - pure $C_{60}$.

On heating, the thermal expansion of the 9.9% $N_2 – C_{60}$ sample exhibited negative and positive contributions with different characteristic times of relaxation. The negative contribution was only observed in the interval 4.5 – 8 K. It was not detected, within the experimental error, at higher and lower temperatures. The negative contribution to the thermal expansion was absent when the sample was cooled. Much similar behavior of the thermal expansion coefficient was observed earlier in $C_{60}$ doped with $H_2$, $D_2$, Kr and Xe [3,8].

The experiments with high $N_2$ concentrations were made on a sample of high-purity (99.99%) $C_{60}$ powder (SES, USA) with the average grain size of about 100 μm.

The $C_{60}$ powder was saturated with $N_2$ under ~ 200 MPa at T=575°C for 36 hour at the Australian Nuclear Science and Technology Organization (Australia). The TGA (thermal gravimetric analysis) results show that the octahedral cavities were practically completely (100%) occupied by $N_2$. The $N_2$-saturated powder was compacted at Umea University, Sweden, using the technique described in [8]. The final sample was a cylinder 8 mm high and 10 mm in diameter.

The sample was delivered to Ukraine by mail. On its way, it was broken into two unequal parts. The investigation was made on the larger part which was worked mechanically into a 4.1 mm high cylinder.

The cylindrical sample was placed into the measuring cell of the dilatometer and cooled to liquid helium temperature by the technique used for $C_{60}$ with low $N_2$ concentrations (see above).

The obtained temperature dependence of the linear thermal expansion coefficient of $C_{60}$ containing a high $N_2$ concentration (100%) is shown in Fig. 2. The curves shown are averaged over several series of measurement.

At a high $N_2$ concentration the hysteresis in α(T) is much larger than in the case of a low $N_2$ concentration. On heating the thermal expansion of this sample also exhibited negative and positive contributions. However, the negative contribution was observed in a broader interval (2.8 – 20K). At low $N_2$ concentrations it was 4,55 – 8K (see Fig.1). The negative contribution was not found on cooling $C_{60}$ with a high $N_2$ concentration. It is interesting to consider some features of the α(T) behavior observed in the $C_{60}$ sample with a high $N_2$ concentration.

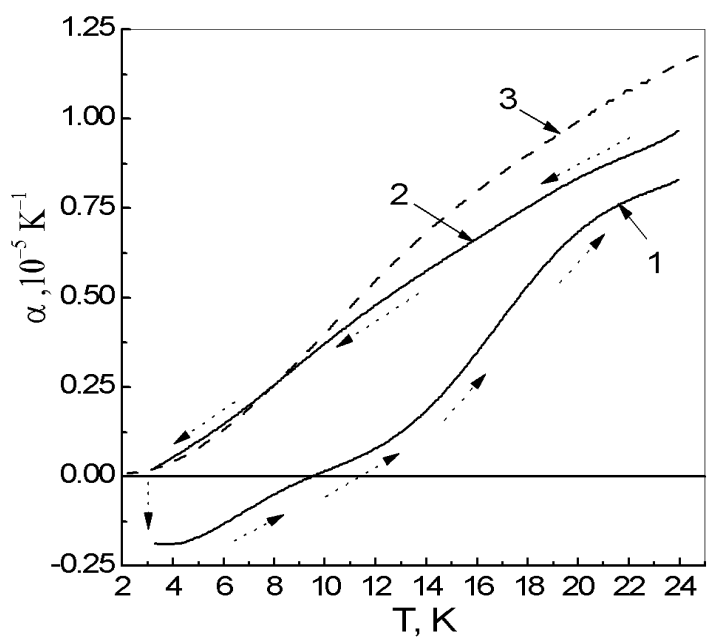

Fig.2 . Temperature dependence of the linear thermal coefficient of pure $C_{60}$ and $C_{60}$ with a high $N_2$ concentration. 1– $N_2$-$C_{60}$ on heating, 2 – $N_2$-$C_{60}$ on cooling. 3 – pure $C_{60}$ on heating and cooling.

1. To change from the condition of curve 2 (cooling to the lowest temperature of the experiment T=2.8K) to subsequent heating when the dependence α(T) is described by curve 1, the sample was kept at T=2.8K at least for 2.5 hours.
2. If the sample is not warmed above 4.6 K, the negative contribution is observable both on the heating and cooling of the sample. However, after heating the sample above 6.5 K, the negative contribution disappears when the sample is cooled.

**Discussion**.

The hysteresis in the dependence α(T) observed in gas – fullerite solutions is caused by the co-existence and mutual transformations of different orientational glasses $N_2$-$C_{60}$ [3,8,13] .

In solutions with high and low $N_2$ concentrations the low temperature glass phase (phase I) is in equilibrium at least in the interval 2.8 – 4.6K. For $C_{60}$ with a high $N_2$ concentration this is evident from the fact that after the first – order phase transition at T=2.8K the cycling of the sample within the interval 2.8 – 4.6K does not cause a hysteresis in α(T). The dependence α(T) of $C_{60}$ with a low $N_2$ concentration has no hysteresis in this temperature interval. The dependence α(T) of phase I is described in Figs. 1 and 2 by curves 1.

In solutions with high and low $N_2$ concentrations, the "high temperature" phase ( phase II) is in equilibrium at least in the interval 6-24K. The function α(T) for phase II is described in Figs. 1 and 2 by curves 2. There is certain evidence for the equilibrium state of phase II. The α – values that fall on curves 2 in this temperature interval still hold their positions after sign – alternating variations of temperature by 1-3K (cycling). As for the α – values on curves 1, several cycles are sufficient to shift them to the corresponding curves 2. The reproducibility of the measured data was poor in the interval 4.6 – 6K. This region of instability may suggest the presence of two phases here. We can assume that this interval includes the temperature of the phase transition, which corresponds well with theoretical estimates [8].

The derived information can be used to specify the features of phases I and II. As was mentioned above, on heating the sample by ΔT, the time dependence of the thermal expansion exhibits two processes with different characteristic times.

At constant temperature the dependence α(T) measured on cooling can be described by

$$\alpha(t) = \frac{1}{\Delta T} \cdot \frac{\Delta L}{L} = A(1-exp(-t/\tau_1)) + B(exp(-t/\tau_2) - 1) ,  \qquad (2)$$

where the first and second terms on the equation's right-hand side describe the positive and negative contributions respectively; $A$ and $B$ are the absolute values of the corresponding contributions at $t \to \infty$; and $\tau_1$ and $\tau_2$ are the characteristic relaxation times for these contributions.
Using the data processing procedure of [3], we evaluated the positive (A) and negative (B) contributions and the characteristic times of the processes responsible for the thermal expansion of the $N_2$-$C_{60}$ samples as a function of temperature.

It appears that the positive contribution α(T) to the thermal expansion of phase I coincides within the experimental error with the thermal expansion α(T) in phase II. Thus, the low frequency excitations ( phonons and librons ) contribute equally to the thermal expansions in phases I and II. This is quite natural if we recall that there is only a minor distinction between the molar volumes of the phases [14]. At the same time the processes of $C_{60}$ molecule reorientation which determine the negative contribution to the thermal expansion manifest themselves only in phase I.

The temperature dependences of the ratio between the negative (B) and positive (A) contributions to the thermal expansion of phase I with low and high concentrations of nitrogen are shown in Fig.3a.

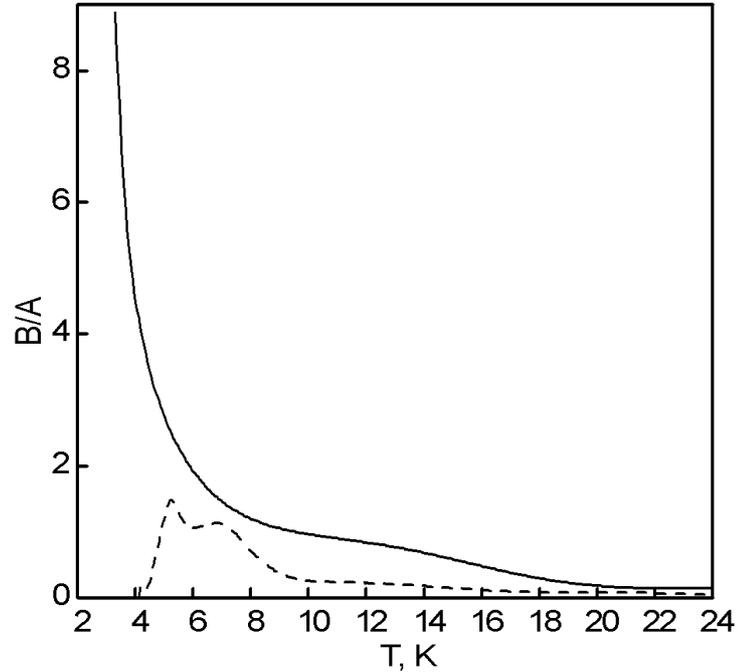

Fig.3a. The absolute value of the ratio between negative and positive contributions to the thermal expansion of $C_{60}$ sample with a high ( solid line) and low and low ( broken line) $N_2$ concentration.

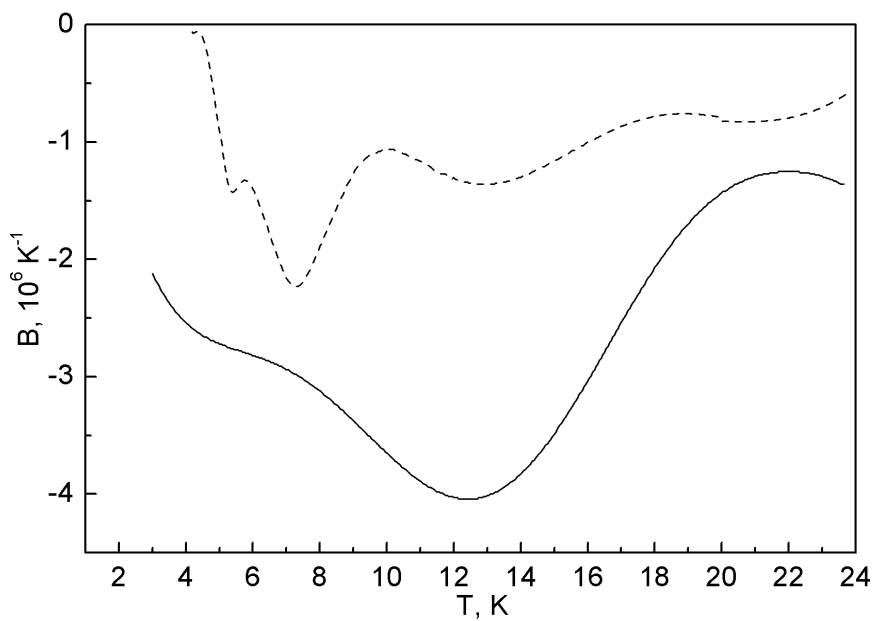

Fig.3b. The negative contribution to the thermal expansion of $C_{60}$ sample with a high (solid line) and low (broken line) $N_2$ concentration.

Fig. 3b illustrates the dependence of the negative contributions (B) to the thermal expansion of $C_{60}$ samples with high and low $N_2$ concentrations. It is seen that the negative contribution (B) and the interval of its existence increase when the $N_2$ concentration grows.

We can assume that at growing $N_2$ concentrations more and more $C_{60}$ molecules are involved in the process of reorientation and the potential barriers impeding reorientation become suppressed. The first factor is responsible for the increase in B. The other leads to the broadening of the band of the tunnel $C_{60}$ rotation spectrum and hence to extension of the temperature interval where the negative thermal expansion is observable.

As was found in [3,8], the negative contribution to the thermal expansion of $C_{60}$ is largely determined by the diameter of the dissolved molecules rather than by their mass. It is therefore interesting to compare the dependence B(T) of the $N_2$-$C_{60}$ solution with the corresponding dependence for $C_{60}$ containing equal concentrations of other gas. It is convenient to start with a solution having a low content of impurity, where the interference of the interaction between the impurity particles is negligible. In our previous investigations [1-3, 6-8] low concentrations of the dissolved gases were undetectable. In this study we tried the technique [8] used to saturate pure $C_{60}$ with deuterium and then measured the $D_2$ concentration. It was 12.9%, i.e. very close of the $N_2$ concentrations measured in this study. This justifies our comparison of B(T) measured in the $N_2$-$C_{60}$ and $D_2$-$C_{60}$ solutions. We see in Fig.4 that the dependence B(T) of the $N_2$-$C_{60}$ solution resembles that of the $D_2$-$C_{60}$ solution but is shifted towards high temperatures and stretched along the T-axis. The explanation can be as follows. The introduced $N_2$ molecules push apart the neighboring $C_{60}$ molecules more vigorously than the smaller $D_2$ molecules can do. The distance between the $C_{60}$ molecules increases and their non central interaction attenuates, which suppresses the barriers impeding the reorientation of the $C_{60}$ molecules. This in turn broadens the band of the tunnel rotation spectrum of $C_{60}$ molecules. As a result, the dependence B(T) appears to be shifted towards high temperatures and stretched along the T-axis. Meanwhile, in both $N_2$-$C_{60}$ and $D_2$-$C_{60}$ solutions the behavior of B(T) remains practically unaffected by the rather large difference between the masses of $N_2$ and $D_2$ molecules.

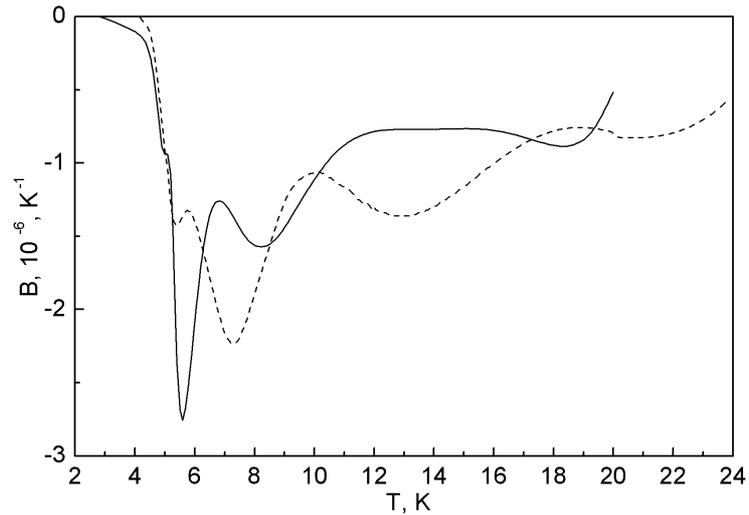

Fig.4. Absolute values of negative contributions B to the linear thermal expansion coefficient of $C_{60}$+ 9,9% $N_2$ (solid line) and $C_{60}$+12.9% $D_2$(broken line).

Fig.5 shows the temperature dependences of the characteristic times of the positive ($\tau_1$) and negative ($\tau_2$) contributions to the thermal expansion of $C_{60}$ with low and high $N_2$ concentrations. $\tau_1$ is actually the characteristic time during which the temperature equalizes over the sample. $\tau_1$ is sensitive to the size and thermal resistance of the sample. The influence of gas impurities upon the thermal resistance of $C_{60}$ is rather weak against the background of the high thermal resistance of compressed pure samples of $C_{60}$ [3,8]. The higher $\tau_1$ – values obtained for $C_{60}$ with a low $N_2$ content may be due to the large size of the sample (9mm). The size of the sample with a high $N_2$ concentration was 4mm.

We did not observe any significant dependence of $\tau_2$ on the concentration of the gas impurity in [3,8], which can be due to the much smaller difference between the gas concentrations in the samples. Nevertheless, in this study there were only minor distinctions between the $\tau_2$ – values of the samples with 9,9% $N_2$ and 100% $N_2$ (see Fig.5).

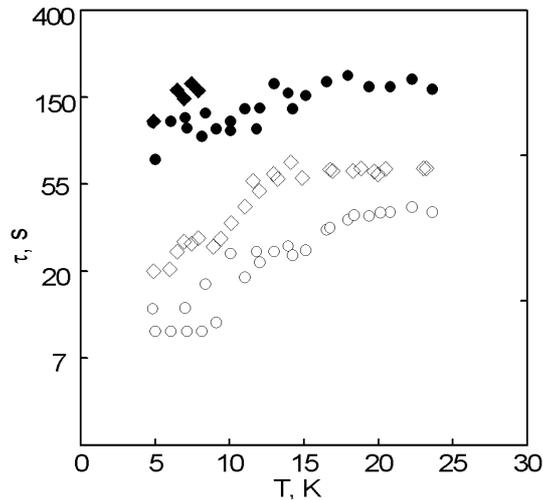

Fig.5. Characteristic times of positive and negative contributions to the thermal expansion of $N_2$-$C_{60}$ sample.
Positive contributions:
◊ - $C_{60}$ with low $N_2$ concentration;
○ – $C_{60}$ with high $N_2$ concentration;
Negative contributions:
♦ - $C_{60}$ with low $N_2$ concentration;
● – $C_{60}$ with high $N_2$ concentration.

$\tau_2$ is the characteristic time of reorientation of $C_{60}$ molecules. It is rather strongly dependent on the type of gas impurity [3,8]. Fig.6 illustrates the characteristic times $\tau_2$ for $C_{60}$ with high and low concentrations of $N_2$ and 12,9% $D_2$. The comparison of the negative contributions B(T) to the thermal expansion coefficients of the $N_2$-$C_{60}$ and $D_2$-$C_{60}$ solutions shows that because of the larger sizes of $N_2$ molecules the probability of $C_{60}$ reorientation is higher in the $N_2$-$C_{60}$ solution. The lower $\tau_2$ – values in the $N_2$-$C_{60}$ sample in comparison to those in the $D_2$-$C_{60}$ system reflect the higher probability of $C_{60}$ reorientation in the $N_2$-$C_{60}$ solution.

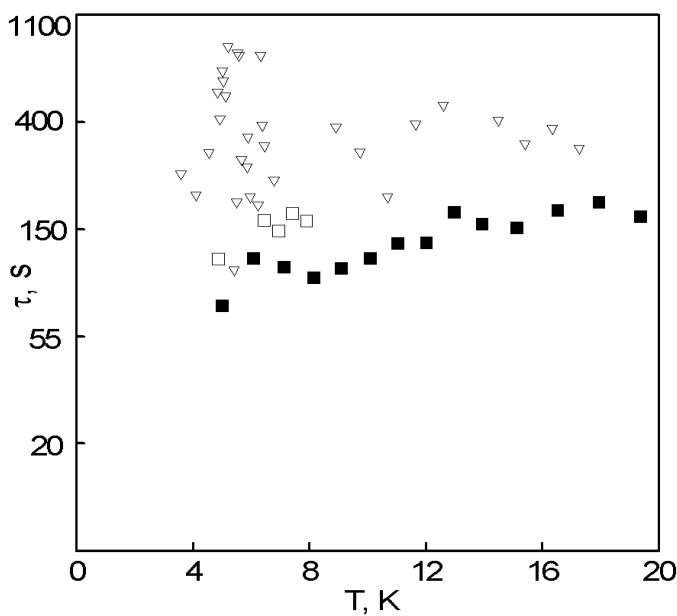

Fig.6. Characteristic times of negative contributions to the thermal expansion of $C_{60}$ doped with $N_2$ and $D_2$:
□ – $C_{60}$ with low $N_2$ concentration;
■ – $C_{60}$ with high $N_2$ concentration;
∇ – $D_2$-$C_{60}$.

We also investigated the kinetics of phase transformations in the orientational glasses $N_2$-$C_{60}$. The characteristic time $\tau'$ of the phase transition was found through sign – alternating variations of temperature by 1-3K (thermocycling) in the nonequilibrium phase. During thermocycling the $\alpha$ – value varied approaching the thermal expansion coefficient of the equilibrium phase. The characteristic time $\tau'$ was obtained from the time dependence $\alpha(t)$. The experimental technique and data processing are detailed elsewhere [8]. Fig. 7 shows $\tau'$ versus the average temperature of thermocycling. In the whole interval of temperatures the results on $\tau'$ obtained for $C_{60}$ with a low $N_2$ concentration are reliable to within the experimental error. The dependence $\tau'(T)$ taken on $C_{60}$ with high $N_2$ concentration has a maximum at T ≈ 11.5K. $\tau'$ was only slightly dependent on the $N_2$ concentration.

The phases of the orientational glasses of $C_{60}$ saturated with gases differ mainly in the orientational order of the $C_{60}$ molecules [8]. It is therefore interesting to compare the characteristic times $\tau'$ (phase transformation) and $\tau_2$ ($C_{60}$ reorientation). In the $N_2$-$C_{60}$ solution $\tau'$ is about an order of magnitude higher than $\tau_2$. Note that in this solution the correlation between $\tau'$ and $\tau_2$ is observed only at one temperature, T=11.5K [8].

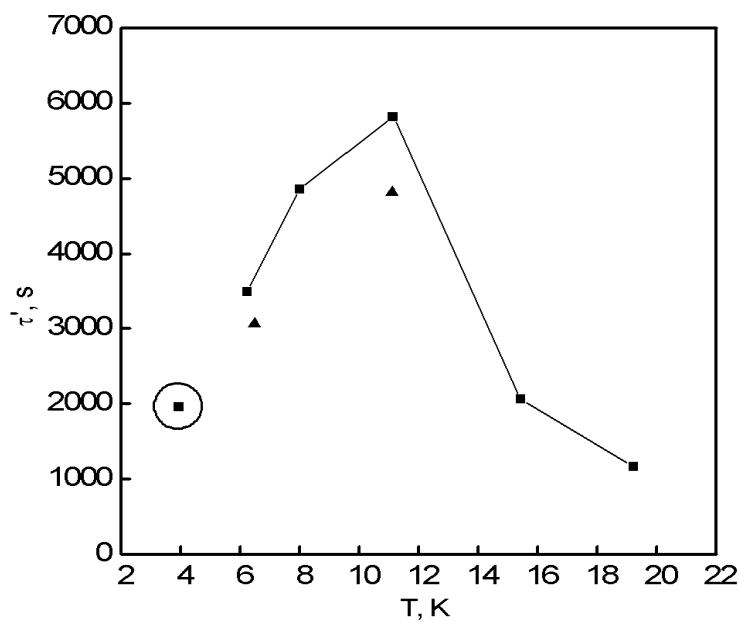

Fig.7. Temperature dependence of characteristic time $\tau'$ of phase transformation in orientational $N_2$-$C_{60}$ glasses.
Change from phase I to phase II:
▲  - $C_{60}$ with low $N_2$ concentration;
■  - $C_{60}$ with high $N_2$ concentration.
Change from phase II to phase I:
Ⓜ  - $C_{60}$ with high $N_2$ concentration.

The above facts suggest the following conclusions.
1. In the temperature interval 2.2 – 24K, fullerite $C_{60}$ doped with gases exhibits coexistence of several types of orientational glasses, each of them having its characteristic molar volume.
2. A change in the temperature condition of sample leads to mutual phase transformations of tunnel character. We have estimated the temperature dependence of the characteristic time of such transformations (Fig.7).
3. For the systems investigated there is a certain limiting temperature $T_c$ below which one of the orientational glass phases becomes the main contributor to the thermal expansion. This is $T_c \approx 4.5K$ for the $N_2$-$C_{60}$ systems investigated in this study.

We wish to thank Prof. A.S Bakai for helpful participation in the discussion of the results.
The authors are indebted to the Science and Technology Center in Ukraine for support.

**References.**

1. A.N. Aleksandrovskii, V.B. Esel'son, V.G. Manzhelii, A. Soldatov, B. Sundqvist, and B.G. Udovidchenko, *Fiz. Nizk.Temp.***23**, 1256 (1997) [*Low Temp. Phys.* **23**, 943 (1997)].

2. A.N. Aleksandrovskii, V.B. Esel'son, V.G. Manzhelii, A. Soldatov, B. Sundqvist, and B.G. Udovidchenko, *Fiz. Nizk.Temp.* **26**, 100 (2000) [*Low Temp. Phys.* **26**, 75 (2000)].

3. A.N. Aleksandrovskii, A.S Bakai, A.V. Dolbin, V.B. Esel'son, G.E. Gadd, V.G. Gavrilko, V.G. Manzhelii, S. Moricca, B. Sundqvist, B.G. Udovidchenko, Fiz. Nizk. Temp. **29,** 432-442, (2003) (Low Temp. Phys. **29**, 324-332, (2003)).

4. V. M. Loktev, J. N. Khalack, Yu. G. Pogorelov, *Fiz. Nizk. Temp.* **27,** 539, (2001)]. [*Low Temp. Phys.* **27,** 397, (2001)].
5. J. M. Khalack and V. M. Loktev, *Fiz. Nizk. Temp.* **29,** 577 (2003) [*Low Temp. Phys* **29,** 429 (2003)].
6. A.N. Aleksandrovskii, V.G. Gavrilko V.B. Esel'son, V.G. Manzhelii, B. Sundqvist B.G. Udovidchenko, and V.P. Maletskiy, *Fiz. Nizk. Temp.* **27**, 333 (2001) [*Low Temp. Phys.* **27**, 245 (2001)]
7. A.N. Aleksandrovskii, V.B. Esel'son, V.G. Gavrilko, V.G. Manzhelii, B. Sundqvist, B.G. Udovidchenko and V.P. Maletskyi, *Fiz. Nizk.Temp.* **27**, 1401 (2001) [*Low Temp. Phys.* **27**, 1033 (2001)]
8. A.N. Aleksandrovskii, A.S. Bakai, D. Cassidy, A.V. Dolbin, V.B. Esel`son, G.E. Gadd, V.G. Gavrilko, V.G. Manzhelii, S. Moricca, and B. Sundqvist, *Fiz. Nizk. Temp.* **31**, 565 (2005) [*Low Temp. Phys*. **31**, 429 (2005)].
9. I. V. Legchenkova, A. I. Prokhvatilov, Yu. E. Stetsenko, M. A. Strzhemechny, K. A. Yagotintsev, A. A. Avdeenko, V. V. Eremenko, P. V. Zinoviev, V. N. Zoryanski, N. B. Silaeva, and R. S. Ruoff, *Fiz. Nizk.Temp.* **28**, 1320 (2002) (*Low Temp. Phys.* **28,** 942 (2002)).
10. Yu.E. Stetsenko, I.V. Legchenkova, K.A. Yagotintsev, A.I. Prokhvatilov, and M.A. Strzhemechny, *Fiz. Nizk.Temp.* **29**, 597 (2003) (*Low Temp. Phys.* **29,** 445 (2003)).
11. L.S. Palatnik, A.A. Nechitailo, A.A. Koz'ma, Dokl. Akad. Nauk SSSR, **36**, 1134 (1981).
12. C.A. Angell, Science 267, 1924(1995).
13. A.I. Prokhvatilov, N.N. Galtsov, I.V. Legchenkova, M. A. Strzhemechny, D. Cassidy, G.E. Gadd, S. Moricca, B. Sundqvist, N.A. Aksenova , **31**, 585 (2005).
14. A.I. Prokhvatilov, the private message.